\documentclass[useAMS,referee]{mn2e} 
\usepackage{graphicx}
\def\ee{\end{equation}}
\def\be{\begin{equation}}
\def\bdm{\begin{displaymath}}
\def\edm{\end{displaymath}}

\usepackage{graphics}

\def\be {\begin{equation}}
\def\ee {\end{equation}}
\def\ebb{\end{displaymath}\begin{equation}}

\def\noi{\noindent }

\newcommand{\nup}[1]{\nu^{\prime #1}}
\newcommand{\Dp}{f^{\prime}}
\newcommand{\ergcm}{\;\;\hbox{erg}\;\hbox{cm}^{-2}}
\newcommand{\nus}{\nu_{syn}}
\newcommand{\nut}{\nu_{T}}
\newcommand{\nub}{\nu_{B}}
\newcommand{\nug}{\nu_{\gamma_0}}
\newcommand{\nuc}{\nu_{c}}
\newcommand{\nuta}{\nu_{T}/\alpha^4}
\newcommand{\nuga}{\nu_{\gamma_0}/\alpha}
\newcommand{\us}{\;\hbox{Hz}}
\newcommand{\Rpm}[1]{R^{\prime}_{max}(#1)}
\newcommand{\Rpp}[1]{R^{\prime}_{peak}(#1)}
\newcommand{\Kl}{K\ll 1}
\newcommand{\Kg}{K\gg 1}
\newcommand{\Kla}{1\ll K\ll \alpha^3}
\newcommand{\Kga}{1\ll \alpha^3 \ll K}
\def\sy{synchrotron }

\def\ep {\epsilon }
\def\e0 {\epsilon _0}

\def\noi{\noindent }
\def\a{\alpha }
\title[Ordering parameter of SSC-emitting blazars]{A new ordering parameter of spectral energy distributions from synchrotron-self-Compton emitting blazars}
\author[M. Zacharias \& R. Schlickeiser]{M. Zacharias \& R. Schlickeiser\thanks{E-mail: rsch@tp4.rub.de} \\
Institut f\"ur Theoretische Physik, Lehrstuhl IV: Weltraum- und Astrophysik, Ruhr-Universit\"at Bochum, 
44780 Bochum, Germany}
\date{Received ?; accepted ? }
\begin{document}
\maketitle
\begin{abstract}
The broadband SEDs of blazars exhibit two broad spectral components, which in leptonic emission models are attributed to synchrotron radiation and synchrotron self-Compton (SSC) radiation of relativistic electrons. During high state phases, the high-frequency SSC component often dominates the low-frequency synchrotron component, implying that the inverse Compton SSC losses of electrons are at least equal to or greater than the synchrotron losses of electrons. We calculate from the analytical solution of the kinetic equation of relativistic electrons, subject to the combined { linear} synchrotron and nonlinear synchrotron self-Compton cooling, for monoenergetic injection the time-integrated total \sy and SSC radiation fluences and spectral energy distributions (SED). Depending on the ratio of the initial cooling terms, displayed by the { injection} parameter $\a$, we find for $\a\ll 1$, { implying complete linear cooling,} that the \sy peak dominates the inverse Compton peak and the usual results of the spectra are recovered. For $\a\gg 1$ the SSC peak dominates the \sy peak, proving our assumption that in such a case the cooling becomes initially non-linear. The spectra also show some unique features, which can be attributed directly to the non-linear cooling. To show the potential of the model, we apply it to { outbursts of 3C 279 and 3C 454.3, successfully reproducing the SEDs.} The results of our analysis are promising, and we argue that this non-equilibrium model should be considered in future modeling attempts for blazar { flares}.  
\end{abstract}
%
%
\section{Introduction}
{ Blazars are the most energetic and most violent subclass of active galaxies (Urry \& Padovani 1995). Their spectral energy distributions (SEDs) extend through all bands and are dominated by two distinct broad nonthermal components. The low-energetic component, extending from radio to optical or even X-ray frequencies, is attributed to synchrotron emission of highly relativistic electrons moving in a compact blob down the jet. The high-energetic component covers the X-ray and $\gamma$-ray bands, and is interpreted in leptonic models as inverse Compton scattered photons of the same electron population (for a review see B\"ottcher 2007). 

Several photon sources have been proposed as seed fields for the inverse Compton process, e.g. from the accretion disk sorrounding the black hole in the center of the active galaxy (Dermer \& Schlickeiser 1993), from the broad line regions (Sikora et al. 1994) or the dusty torus (Blazejowski et al. 2000). Normally, such models are referred to as external, since the photons are produced externally to the jet. As an internal source of photons the synchrotron radiation of the electrons can be used for Compton scattering, as well, which is referred to as the synchrotron-self Compton (SSC) process (Jones, O'Dell \& Stein 1974). 

In many blazars the Compton component dominates the \sy component, while the peak frequencies are rather low (\sy component peaking in the infrared), like in 3C 279 (Abdo et al. 2010) or 3C 454.3 (Vercellone et al. 2011). This is normally interpreted as a result of inverse Compton scattering of photons produced externally to the jet (e.g. Sikora et al. 2009, Bonnoli et al. 2011). Due to the strong radiation fields the cooling of the electrons would become severe, explaining the low peak frequencies. On the other hand, blazars with a less luminous Compton peak, compared to the \sy peak, are successfully modeled with the SSC approach, like 1ES 2344+514 (Acciari et al. 2011). These sources typically have much higher peak frequencies, with the peak of its \sy component even as high as the X-ray regime. These differences between blazars has given rise to the so-called blazar sequence, which was found by Fossati et al. (1998), and interpreted by Ghisellini et al. (1998). However, recently these interpretations have been challenged as being due to selection effects (Gupta et al 2011, Giommi et al. 2011).

As a matter of fact, from a theoretical point of view it is not immediately clear, why a dominance of the inverse Compton peak over the synchrotron peak (also referred to as the Compton dominance) should favour external photons as the main contributer to electron cooling over the SSC process. Assuming that the electron cooling works mainly via the \sy and SSC channels, then the ratio of the observed SSC to \sy photon luminosities from the same population of electrons $n(\gamma )$ becomes
\be
{L^*_{SSC}\over L^*_{sy}}={\int dV\int_1^\infty d\gamma \; n(\gamma )\vert \dot{\gamma _{SSC}}\vert \over 
\int dV\int_1^\infty d\gamma \; n(\gamma )\vert \dot{\gamma _S}\vert }.
\label{a0}
\ee
Therefore, this directly reflects the ratio of the corresponding loss rates, because of the identical Doppler boosting factors of \sy and SSC emission (Dermer and Schlickeiser 2002). Hence, a large Compton dominance implies in this case that the electrons should mainly cool by SSC cooling, instead of \sy cooling.

It was pointed out by Schlickeiser (2009, hereafter referred to as paper S) that the SSC cooling rate becomes in the Thomson-limit
\be
|\dot{\gamma }|_{\rm SST}= A_0\gamma ^2\int_0^\infty d\tilde\gamma \, \tilde\gamma^2 n(\tilde\gamma ,t) \;,
\;\; 
A_0={3\sigma _Tc_1P_0R\e0 ^2\over mc^2},
\label{a1}
\ee
where $\gamma$ is the electron Lorentz factor, $P_0=3.2\cdot 10^{12}$ eV$^{-1}$s$^{-1}$, $\e0 =1.16\cdot 10^{-8}b$ eV (and $b$ reflects the strength of the magnetic field $B=b$ Gauss), $R$ is the radius of the spherical source, $m$ the electron rest mass, $c$ denotes the speed of light, $\sigma _T=6.65\cdot 10^{-25}$ cm$^2$ is the Thomson cross section and $c_1=0.684$.

Comparing Eq. (\ref{a1}) to the normal \sy cooling term
\be
|\dot{\gamma }|_S=D_0\gamma ^2,\;\;\; D_0={4\over 3}{c\sigma _T\over mc^2}U_B=1.29\cdot 10^{-9}b^2\;\,\, \hbox{s}^{-1}
\label{a2}
\ee
(Blumenthal \& Gould 1970), with $U_B=B^2/8\pi$, one finds that the SSC cooling term depends on an integral over the electron distribution itself, making the cooling term nonlinear and time-dependent. This has remarkable consequences. High energetic electrons will be severely cooled, which results in the fact that any broad distribution of electrons will be quenched very fast into a very narrow, almost $\delta$-like distribution. This was discussed in detail by Zacharias \& Schlickeiser (2010, hereafter referred to as ZS). On the other hand, while the electrons lose energy the cooling rate will also become lower. This implies that after some time the cooling will eventually become linear, i.e. the electrons will cool by the \sy term (Eq. (\ref{a2})). This can be accounted for by combing the two cooling terms into a single one, as was done by Schlickeiser, B\"ottcher \& Menzler (2010, hereafter referred to as SBM) and also by ZS:
\be
|\dot{\gamma }|_{TOT}=|\dot{\gamma }|_S + |\dot{\gamma }|_{\rm SST} = \gamma ^2\left(D_0+A_0\int_0^\infty d\gamma \, \gamma^2 n(\gamma ,t)\right).
\label{a03}
\ee

Another important consequence of the time-dependency of the cooling term is that the resulting electron distribution will not be in an equilibrium state. Such states are mostly invoked to calculate the underlying electron distribution (see e.g. Abdo et al. 2011, Aleksic et al. 2011). However, in order to account for the characteristic spectral breaks and features in blazar SEDs such equilibrium calculations need to invoke breaks in the electron distribution (e.g. Sanchez et al. 2010) with little physical justification. 

There are several attempts in the literature to calculate blazar spectra with combined cooling scenarios. Moderski et al. (2005) analysed cooling under synchrotron and inverse Compton losses, taking into account Klein-Nishina-effects on the inverse Compton cooling term. However, their analyses did not treat a time-dependent cooling term and, thus, they also conducted their calculations for the steady-state scenario. A similar approach is used also by Nakar, Ando \& Sari (2009). As B\"ottcher \& Dermer (2010) pointed out, a full time-dependent treatment including Klein-Nishina effects can mostly be solved only numerically. Such numerical studies have been performed by, e.g., Li \& Kusunose (2000), or Kato, Kusunose \& Takahara (2006). Other numeical works, like Katarzynski et al. (2006), also include acceleration processes, giving probably a more complete picture. Although this list is, of course, incomplete, to our knowledge no one has tried so far calculating analytically the complete spectrum of blazars in a time-dependent scenario, which includes both the SST and \sy cooling terms given in Eq. (\ref{a03}).

This is what we attempt in the present work. The strategy would be to solve the kinetic equation for the electron distribution, calculate the resulting synchrotron spectrum, and then the SSC spectrum. However, we can take advantage of the work of SBM, who already performed the first and the second step. Hence, we are left here with the task to calculate the SSC spectrum.

Since we rely on the work of SBM, we will also use their approach of a monoenergetic instantaneous injection of electrons into the blob. Thus, our calculations can be regarded as a description of a single outburst in a blazar. The monoenergetic approach can be justified in two ways. Firstly, nearly monoenergetic electron injection distributions result from the pile-up mechanism (Schlickeiser 1984, Schneider 1993, Jauch and Duschl 1999), i.e. the simultaneous operation of first-order Fermi acceleration and radiative losses of electrons, and the pick-up mechanism (Pohl and Schlickeiser 2000). Henri \& Sauge (2006) also argued in favor of an electron distribution close to a monoenergetic one in blazars, at least in case of a more complicated stratified jet structure. Secondly, cooling is most effective for high energetic electrons. As was shown by ZS for a power-law ($n(\gamma)\propto \gamma^{-s}$), and already mentioned above, cooling leads to a rapid quenching of any broad distribution, resulting also in a near monoenergetic one. This quenching operates even faster for the non-linear cooling. Thus, the monoenergetic approach is at least valid in a late time limit. This can also be seen if one compares the resulting synchrotron spectra in SBM and ZS. Several of the characteristic features are the same regardless of the approach. They only differ at the high-energy end of the synchrotron spectrum. Instead of an exponential cut-off produced by a monoenergetic distribution, the power-law results in just another power-law depending on the spectral index $s$ of the electrons. Therefore, only the high-energetic part of the \sy spectrum actually depends on the injected form of the distribution. Anything below a certain frequency comes from a distribution that is already almost monoenergetic. Finally, using the monoenergetic approach also puts us in a position to calculate the SSC spectrum at all. Analytical calculations with broad distributions are at least difficult (e.g. Dermer, Sturner \& Schlickeiser 1997) if not impossible.

Our main intention of this paper is to demonstrate the implications of the combined cooling term. Therefore, apart from the non-linear cooling term and the full time-dependent treatment, we will keep the calculations simple. The optically thick case, which is important only for very low photon energies, as well as photo-pair-production in the jet or external by the extragalactic background light are not treated here. Since both are probably important due to the characteristics of the source for the resulting spectra, we will consider them in future work. We also neglect the effect of non-radiative adiabatic cooling, which is due to the finite opening angle of the jet. 

In order to give a complete picture of the problem, we review and summarise the results of SBM in section 2 and 3. There we discuss the solution for the kinetic equation of the electron distribution and the emerging \sy spectra. We also focus our attention on the definition and interpretation of the injection parameter $\alpha$, which is also the new ordering parameter. In section 4 we then calculate the SSC spectra. Readers who are less interested in the mathematical details may skip this section, since we sum up the results in section 5. In this section we give the theoretical spectra in the form of SEDs transformed into the system of rest of the host galaxy, which is for low redshift more or less the observers frame. There we also calculate the values of the Compton dominance, as well as the ratio of the peak frequencies. In section 6 we apply our model to flares of the blazars 3C 279 and 3C 454.3, and discuss the resulting spectra. Finally, we summarise and conclude in section 7.}
%
%
\section{Solutions of the electron kinetic equation for combined synchrotron and SST cooling}
The competition between the instantaneous injection of ultrarelativistic electrons ($\gamma \gg 1$) with the arbitrary injection distribution function  $q(\gamma )$ at time $t=0$ and the combined synchrotron and SST radiative losses (\ref{a03}) is described by the time-dependent kinetic equation for the volume-averaged relativistic electron population inside the radiating source (Kardashev 1962):  

\be
{\partial n(\gamma ,t)\over \partial t} 
-{\partial \over \partial \gamma }\left[\gamma ^2\left(D_0+A_0\int_0^\infty d\gamma \, \gamma^2 n(\gamma ,t)\right)n(\gamma ,t)\right]=q(\gamma )\delta (t)
\label{a3}
\ee
where $n(\gamma ,t)$ denotes the volume-averaged differential electron number density.

The kinetic equation (\ref{a3}) has been solved for monoenergetic electron injection $q(\gamma )=Q_0\delta (\gamma -\gamma _0)$ by SBM. They demonstrated that the solutions of the kinetic equation (\ref{a3}) sensitively depend on the injection parameter 

\be
\a =\frac{|\dot{\gamma }(t=0)|_{\rm SST}}{|\dot{\gamma }|_S}=\gamma_0\left( \frac{Q_0A_0}{D_0} \right)^{1/2} \equiv {\gamma _0\over \gamma_B}=46{\gamma _4N_{50}^{1/2}\over R_{15}}.
\label{a4}
\ee
{ This is our new ordering parameter, and it is defined as the ratio of the nonlinear cooling term to the linear cooling at time of injection ($t=0$). It, therefore, reflects the initial conditions of the source favouring either initial nonlinear ($\a\gg 1$) or linear ($\a\ll 1$) cooling. 

As discussed before, the ratio of the cooling terms is directly related to the ratio of the luminosities of the inverse Compton and \sy peak, c.f. Eq. (\ref{a0}). Hence, the value of $\a$ gives a clear indication, which peak dominates the spectrum. In other words, $\a$ can be inferred from a measured spectrum by the Compton dominance, and also by some other features of the spectrum, as we will discuss later. 

\be
\gamma_B=\left({D_0\over A_0Q_0}\right)^{1/2}={217R_{15}\over N_{50}^{1/2}}
\label{a5}
\ee
is the characteristic injection Lorentz factor reflecting the injection conditions of the relativistic electrons in a spherical homogeneous source of size $R=10^{15}R_{15}$ cm. In Eqs. (\ref{a4}) and (\ref{a5}) we additionally scale $\gamma_0=10^4\gamma _4$ and the total number of injected electrons as $N=10^{50}N_{50}$, so that $Q_0=3N/(4\pi R^3)=2.39\cdot 10^4N_{50}R_{15}^{-3}$ cm$^{-3}$. 

Obviously, a higher density of injected particles leads to a smaller value of $\gamma_B$. This, in turn, favours initial non-linear cooling.}

For small injection parameters $\a \ll 1$ SBM obtained the solution 

\be
n(\gamma ,t,\a \ll 1)\simeq Q_0H[\gamma _0-\gamma ]\delta \left(\gamma -{\gamma _0\over 1+D_0\gamma _0t}\right),
\label{a6}
\ee
which is solely determined by the linear \sy losses (\ref{a2}). 

For large values of the injection parameter $\a \gg 1$ SBM found at early times  

\be 
0\le t\le t_c={\a ^3-1\over 3\gamma_BD_0\a ^3}\simeq {1\over 3\gamma_BD_0}={1.2\cdot 10^6N_{50}^{1/2}\over R_{15}b^2}\;\;\; \hbox{s}
\label{a7}
\ee
the solution 

\bdm
n(\gamma ,\gamma _0,t\le t_c,\a \gg 1)\simeq Q_0H[\gamma _0-\gamma ]H[t_c-t]\delta \left(\gamma -{\gamma _0\over (1+3D_0\gamma _0\a ^2t)^{1/3}}\right)
\edm
\be
=Q_0H[\gamma _0-\gamma ]H[t_c-t]\delta \left(\gamma -{\gamma _0\over (1+3A_0Q_0\gamma _0^3t)^{1/3}}\right),
\label{a8}
\ee
which agrees with the nonlinear SST solution { Eq. (S-28) of paper S.} 

For late times $t\ge t_c$ SBM obtained 

\be
n(\gamma ,\gamma _0,t\ge t_c,\a \gg 1)\simeq Q_0H[\gamma_B-\gamma ]H[t-t_c]\delta \left(\gamma -{\gamma_B\over {1+2\a ^3\over 3\a ^3}+D_0\gamma_Bt}\right), 
\label{a9}
\ee
a modified linear cooling solutions. Note that both solutions (\ref{a8}) and (\ref{a9}) indicate that at time $t_c$ the electrons have cooled to the characteristic Lorentz factor $\gamma_B$, { and, thus, at this point the source turns from non-linear to linear cooling.} 
%
%
\section{Synchrotron total fluence distributions for combined synchrotron and SST cooling}
With the known electron distribution functions (\ref{a6}), (\ref{a8}) and (\ref{a9}) SBM calculated the optically thin \sy radiation intensities 
{
\be
I_S(\ep ,t)=Rj_S(\ep ,t)=\frac{R}{4\pi}\int_0^{\infty}d\gamma\, n(\gamma,t)p_S(\ep ,\gamma) ,
\label{b01}
\ee
where 
\be
p_S(\ep, \gamma)=\frac{P_0\ep}{\gamma^2} CS\left( \frac{\ep}{E_0} \right)
\label{b02}
\ee
denotes the synchrotron power of a single electron in a large-scale randomly oriented magnetic field of constant strength $B$ (Crusius \& Schlickeiser 1988), with the initial characteristic \sy photon energy $E_0={3\over 2}\epsilon_0\gamma_0^2=1.74b\gamma _4^2$ eV.} 

In order to collect enough photons, intensities are often averaged or integrated over long enough time intervals. For rapidly varying photon intensities this corresponds to fractional fluences which are given by the time-integrated intensities. The total fluence spectra are $F(\ep )=\int_0^\infty dt\, I(\ep ,t)$. 

{ For convenience,} we introduce the dimensionless photon energies 
\be
k=\ep /mc^2,\,\; k_0=E_0/mc^2=3.4\cdot 10^{-6}b\gamma _4^2.
\label{b1}
\ee
For small injection energy $\a\ll 1$ the total \sy fluence according to SBM is  
\be
F_s(k,\a \ll 1)\simeq {F_0RQ_0k_0\over D_0\gamma _0^3}({k\over k_0})^{-1/2}\int_{k/k_0}^\infty dy\, y^{-1/6}e^{-y}
\simeq {F_0Rc_0Q_0k_0\over D_0\gamma _0^3}({k\over k_0})^{-1/2}e^{-k/k_0} ,
\label{b2}
\ee
with the constants 
\be
F_0={a_0P_0mc^2\over 8\pi }=7.5\cdot 10^{16}\;\;\; \hbox{s}^{-1},
\label{b3}
\ee
$c_0=0.95302$, and $a_0=1.151275$. 

For the high injection energy case $\a \gg 1$ the total \sy fluence varies as  
\bdm
F_s(k,\a \gg 1)\simeq 
{F_0RQ_0k_0\over D_0\gamma _0^3}({k\over k_0})^{-1/2}\left[{1\over \a ^2}{k_0\over k}\int_{k/k_0}^{k\a ^2/k_0} dy\, y^{5/6}e^{-y}+\, 
\int_{k\a ^2/k_0}^\infty dy\, y^{-1/6}e^{-y}\right]
\edm
\be
\simeq 
{F_0c_0RQ_0k_0\over D_0\gamma _0^3}({k\over k_0})^{-1/2}{k_c\over k+k_c}e^{-k/k_0},
\label{b4}
\ee
with the characteristic \sy peak energy 

\be
k_c={0.703k_0\over \a ^2}=1.13\cdot 10^{-9}{bR_{15}^2\over N_{50}}
\label{b5}
\ee
The analytically calculated \sy SEDs are given by $kF(k)$ and agree well with the corresponding numerically calculated quantities using the radiation code of B\"ottcher et al. (1997). Based on their analysis SBM proposed the following interpretation of multiwavelength blazar SEDs: 

Blazars, where the $\gamma $-ray fluence is much larger than the \sy fluence, are regarded as high 
injection energy sources. Here, the \sy fluence should exhibit the symmetric broken power law behaviour of 
{ Eq. (\ref{b4})} around the \sy peak energy $k_c$. { We note that the injection parameter directly influences the range of the second power-law between the break $k_c$ and the cut-off $k_0$, which gives a first indication how to determine $\a$ from the spectrum.} 

Blazars, where the $\gamma $-ray fluence is much smaller than the \sy fluence, are regarded as small injection 
energy sources. Here, the \sy fluence exhibits the single power law behaviour of { Eq. (\ref{b2})} up to a higher \sy peak energy. 

{ With this interpretation we end the summary of the results of SBM, and turn our attention to the calculation of the SSC peak.}
%
%
\section{SSC total fluence distributions for combined synchrotron and SST cooling}
With the known electron distribution functions (\ref{a6}), (\ref{a8}) and (\ref{a9}) the optically thin SSC radiation intensities 
$I_{SSC}(\ep_s ,t)$ and SSC total fluence distributions $F_{SSC}(\ep_s )=\int_0^\infty dt\, I_{SSC}(\ep_s ,t)$ can be calculated. { The SSC intensity is given by

\be
I_{SSC}(\ep_s ,t)=\frac{R}{4\pi}\int_0^{\infty} d\gamma\, n(\gamma,t)p_{SSC}(\ep_s,\gamma),
\label{c0}
\ee
where $\ep_s$ denotes the energy of the scattered photon, and

\be
p_{SSC}(\ep_s,\gamma)=c\ep_s \int_0^{\infty}d\ep\, n_{S}(\ep,t)\sigma(\ep_s,\ep,\gamma)
\label{c01}
\ee 
is the spectral SSC power of a single electron (Schlickeiser 2002, Ch. 4.2). Here we use the differential \sy photon number density

\be
n_{S}(\ep,t) = \frac{4\pi}{c\ep}I_S(\ep,t)
\label{c02},
\ee
which can be calculated with the help of the \sy intensity $I_S(\ep,t)$, and the differential Klein-Nishina cross-section (Blumenthal \& Gould 1970)

\be
\sigma(\ep_s,\ep,\gamma) = \frac{3\sigma_T}{4\ep\gamma^2}G(q,\Gamma)
\label{c03},
\ee
with

\bdm
G(q,\Gamma) = G_0(q)+\frac{\Gamma^2q^2(1-q)}{2(1+\Gamma q)},
\edm
\be
G_0(q) = 2q\ln{q}+(1+2q)(1-q),
\label{c04}
\ee
and

\be
\Gamma = \frac{4\ep\gamma}{mc^2}, \;\; q=\frac{\ep_s}{\Gamma(\gamma mc^2-\ep_s)}
\label{c05}.
\ee}

For the case of only synchrotron cooling (using the electron distribution (\ref{a6})), and the case of only SST cooling (using the electron distribution (\ref{a8}) at all times), this has been done in paper S. 

Although the SSC energy loss formula (\ref{a1}) holds in the Thomson limit, the SSC intensity calculations of paper S used the full Klein-Nishina cross section. Here we repeat these calculations for the case of combined synchrotron and SST cooling, in order to investigate the influence of the ordering parameter $\a $ on the resulting SSC fluence distributions and SSC SEDs. 

As before, we introduce the dimensionless scattered photon energy 

\be
k_s =\frac{\ep_s}{mc^2}
\label{c1}
\ee
and the normalized time scale 

\be
x={t\over t_S},
\label{c2}
\ee
with the \sy half-life time 

\be
t_S={1\over D_0\gamma _0}={7.75\cdot 10^4\over \gamma _4b^2}\;\; \hbox{s}
\label{c3}
\ee
\subsection{SSC intensities}
For small injection parameters $\a \ll 1$ the SSC intensity { was obtained in paper S (Eqs. (S-77) and (S-78)) }

\bdm
I_{\rm SSC}(k_s,x,\a \ll 1)= I_0k_s(1+x)^4H[\gamma _0-k_s(1+x)]H[x]
\edm
\be
\int_0^1 dq \, q^{-1}G_S(q,\gamma _0,k_s,x)CS \left({k_s(1+x)^4\over 4k_0\gamma _0q[\gamma _0-k_s(1+x)]}\right)
\label{c4}
\ee
with 

\be
G_S(q,\gamma _0,k_s,x)=G_0(q)+{k_s^2(1+x)^2(1-q)\over 2\gamma _0[\gamma _0-k_s(1+x)]},
\label{c5}
\ee
and the constant 

\be
I_0={3\sigma _TR^2Q_0^2P_0mc^2\over 16\pi \gamma _0^4}
\label{c52}
\ee

For the high injection energy case $\a \gg 1$ we use Eqs. (S-75) and (S-76) of paper S at early normalized times $0\le x\le x_c$,
where 

\be
x_c={\a ^3-1\over 3\a ^2}\simeq {\a \over 3}
\label{c6},
\ee
giving

\bdm
I_{\rm SSC}(k_s,x\le x_c,\a \gg 1)= I_0k_s(1+3\a ^2x)^{4/3}H[\gamma _0-k_s(1+3\a ^2x)^{1/3}]H[x_c-x]
\edm
\be
\int_0^1 dq \, q^{-1}G_T(q,\gamma _0,k_s,x)CS \left({k_s(1+3\a ^2x)^{4/3}\over 4k_0\gamma _0q[\gamma _0-k_s(1+3\a ^2x)^{1/3}]}\right)
\label{c7}
\ee
with 

\be
G_T(q,\gamma _0,k_s,x)=G_0(q)+{k_s^2(1+3\a ^2x)^{2/3}(1-q)\over 2\gamma _0[\gamma _0-k_s(1+3\a^2x)^{1/3}]}
\label{c8}
\ee
At late times the electron distribution function (\ref{a9}) yields 

\bdm
I_{\rm SSC}(k_s,x\ge x_c,\a \gg 1)= I_0k_s({1+2\a ^3\over 3\a ^2}+x)^4H[\gamma _0-k_s({1+2\a ^3\over 3\a ^2}+x)]H[x-x_c]
\edm
\be
\int_0^1 dq \, q^{-1}G_L(q,\gamma _0,k_s,x)CS \left({k_s({1+2\a ^3\over 3\a ^2}+x)^4\over 4k_0\gamma _0q[\gamma _0-k_s({1+2\a ^3\over 3\a ^2}+x)]}\right)
\label{c9}
\ee
with 

\be
G_L(q,\gamma _0,k_s,x)=G_0(q)+{k_s^2({1+2\a ^3\over 3\a ^2}+x)^2(1-q)\over 2\gamma _0[\gamma _0-k_s({1+2\a ^3\over 3\a ^2}+x)]}
\label{c10}
\ee
In paper S the useful approximation 

\be
J(A)\equiv \int_0^1 dq \, q^{-1}G(q)\, CS \left({A\over q}\right)
\simeq
{63a_0\over 100}A^{-2/3}e^{-A} 
\label{c11}
\ee
was derived { (Eq. (S-91))}. With this approximation we obtain for the SSC intensities (\ref{c4}), (\ref{c7}) and (\ref{c9})

\be
I_{\rm SSC}(k_s,x,\a \ll 1)\simeq {S_0Q_0^2R^2k_0\over \gamma _0^2}H[\gamma _0-k_s(1+x)]({k_s\over k_T})^{1/3}(1+x)^{4/3}\exp [-{k_s\over k_T}(1+x)^4],
\label{c12}
\ee
\bdm
I_{\rm SSC}(k_s,x\le x_c,\a \gg 1)\simeq {S_0Q_0^2R^2k_0\over \gamma _0^2}H[\gamma _0-k_s(1+3\a ^2x)^{1/3}]
\edm
\be
\times
({k_s\over k_T})^{1/3}(1+3\a ^2x)^{4/9}\exp [-{k_s\over k_T}(1+3\a ^2x)^{4\over 3}],
\label{c13}
\ee
and

\bdm
I_{\rm SSC}(k_s,x\ge x_c,\a \gg 1)\simeq {S_0Q_0^2R^2k_0\over \gamma _0^2}H[\gamma _0-k_s({1+2\a ^3\over 3\a ^2}+x)]
\edm
\be
\times
({k_s\over k_T})^{1/3}[{1+2\a ^3\over 3\a ^2}+x]^{4/3}
\exp [-{k_s\over k_T}({1+2\a ^3\over 3\a ^2}+x)^4],
\label{c14}
\ee
respectively, with the Thomson energy 

\be
k_T=4k_0\gamma _0^2=1.36\cdot 10^3b\gamma _4^4
\label{c15}
\ee
and the constant 

\be
S_0={189a_0\over 400\pi }\sigma _TP_0mc^2=1.9\cdot 10^{-7}\;\, \hbox{cm}^2\, \hbox{s}^{-1}
\label{c16}
\ee
\subsection{SSC total fluence}
The total SSC fluencea are given by 

\be
F_{SSC}(k_s)=\int_0^\infty dt\; I_{SSC}(k_s,t)={1\over D_0\gamma _0}\int_0^\infty dx\, I_{SSC}(k_s,x)
\label{d1}
\ee
The fluence behaviour is controlled by { four characteristic normalized energies: }

\noi
(i) the Thomson energy (\ref{c15}), which denotes the maximum initial scattered photon energy in the Thomson limit, 

\noi
(ii) the SR-break energy, introduced first by Schlickeiser and R\"oken (2008),

\be
k_B=\left({\gamma _0^2\over 4k_0}\right)^{1/3}=1.94\cdot 10^4b^{-1/3}
\label{d2},
\ee
which depends weakly ($\propto b^{-1/3}$) on the magnetic field strength,

\noi
(iii) the initial electron injection Lorentz factor $\gamma _0=10^4\gamma _4$, and

\noi
(iv) the characteristic electron Lorentz factor (\ref{a5}), i.e. $\gamma_B=217R_{15}/N_{50}^{-1/2}=\gamma _0/\a $.

Moreover, two different cases have to be considered depending on the value of the { initial} Klein-Nishina parameter 

\be
K=4k_0\gamma _0=0.136b\gamma _4^3
\label{d3}
\ee
being much smaller than unity (initial Thomson limit) or much greater than unity (initial Klein-Nishina limit). 
In the initial Thomson limit we find that $k_T\ll \gamma _0\ll k_B$, whereas in the initial Klein-Nishina limit 
$k_B\ll \gamma _0\ll k_T$. 

\subsection{SSC fluence for small injection case $\a \ll 1$}
In the small injection case we find with Eq. (\ref{c12}) used in Eq. (\ref{d1}) 

\bdm
F_{SSC}(k_s, \a \ll 1)={S_0Q_0^2R^2t_Sk_0\over \gamma _0^2}({k_s\over k_T})^{1/3}H[\gamma _0-k_s]
\int_0^{{\gamma _0\over k_s}-1} dx\, (1+x)^{4/3}\exp [-{k_s\over k_T}(1+x)^4]
\edm
\be
={S_0Q_0^2R^2k_0\over 4D_0\gamma _0^3}({k_s\over k_T})^{-1/4}H[\gamma _0-k_s]\int_{k_s\over k_T}^{({k_B\over k_s})^3} dy\, y^{-5/12}e^{-y},
\label{e0}
\ee
after obvious substitution. 
\subsubsection{Initial Thomson limit}
In the initial Thomson limit $K\ll 1$ the upper integration limit in Eq. (\ref{e0}) $(k_B/k_s)^3\gg 1$ and we have to consider 
the two cases (a) $k_s\ll k_T\ll \gamma _0\ll k_B$ and (b) $k_T\ll k_s\le \gamma _0\ll k_B$, when the lower integration limit is much smaller or greater than unity.
In the first case we obtain approximately 

\bdm
F_{SSC}(k_s\ll k_T, \a \ll 1)\simeq {S_0Q_0^2R^2k_0\over 4D_0\gamma _0^3}({k_s\over k_T})^{-1/4}\int_{k_s\over k_T}^1dy\, y^{-5/12}
\edm
\be
\simeq 
{3S_0Q_0^2R^2k_0\over 7D_0\gamma _0^3}({k_s\over k_T})^{-1/4}[1-(k_s/k_T)^{7/12}],
\label{e1}
\ee
whereas in the second case 

\be
F_{SSC}(k_T\ll k_s, \a \ll 1)\simeq {S_0Q_0^2R^2k_0\over 4D_0\gamma _0^3K^4}({k_s\over k_T})^{-11/3} e^{-k_s/k_T}
\label{e2}
\ee
We combine the last two approximations to 

\be
F_{SSC}(k_s, \a \ll 1,K\ll 1)\simeq {3S_0Q_0^2R^2k_0\over 7D_0\gamma _0^3}({k_s\over k_T})^{-1/4}e^{-k_s/k_T},
\label{e3}
\ee
which agrees with Eq. (S-107) of paper S. 

\subsubsection{Initial Klein-Nishina limit}
In the initial Klein-Nishina limit $K\gg 1$ we consider the two limiting cases (a) $k_s\ll k_B\ll \gamma _0\ll k_T$ and (b) $k_B\ll k_s \le \gamma _0\ll k_T$. 
In the second case both integration limits are small compared to unity so that 

\bdm
F_{SSC}(k_B\ll k_s, \a \ll 1)\simeq {S_0Q_0^2R^2k_0\over 4D_0\gamma _0^3}\int_{k_s\over k_T}^{({k_B\over k_s})^3}dy\, y^{-5/12}
\edm
\be
\simeq 
{3S_0Q_0^2R^2k_0K^{1/3}\over 7D_0\gamma _0^3}({k_s\over k_B})^{-2}[1-(k_s/\gamma _0)^{7/12}]
\label{e4}
\ee
In the first case the upper integration is large compared to unity and we obtain 

\bdm
F_{SSC}(k_s\ll k_B, \a \ll 1)\simeq {3S_0Q_0^2R^2k_0\over 7D_0\gamma _0^3}({k_s\over k_T})^{-1/4}[1-(k_s/k_T)^{7/12}]
\edm
\be
={3S_0Q_0^2R^2k_0K^{1/3}\over 7D_0\gamma _0^3}({k_s\over k_B})^{-1/4}[1-(k_s/k_T)^{7/12}]
\label{e5}
\ee
We combine the last two approximations to 

\be
F_{SSC}(k_s, \a \ll 1,K\gg 1)\simeq {3S_0Q_0^2R^2k_0K^{1/3}\over 7D_0\gamma _0^3}{k^2_B\over ([k_B+k_s]^7k_s)^{1/4}}[1-(k_s/\gamma _0)^{7/12}]
\label{e6}
\ee
In the initial Klein-Nishina limit the synchrotron cooled SSC fluence exhibits a break at the S-R energy $k_B$ from a $\propto k_s^{-1/4}$ to a 
$\propto k_s^{-2}$ power law, confirming the earlier result of Schlickeiser and R\"oken (2008). 

\subsection{SSC fluence for large injection case $\a \gg 1$}
In the large injection case we find with Eqs. (\ref{c13})--(\ref{c14}) used in Eq. (\ref{d1}) 

\be
F_{SSC}(k_s, \a \gg 1)={S_0Q_0^2R^2k_0\over 4D_0\gamma _0^3}H[\gamma _0-k_s]({k_s\over k_T})^{-3/4}
\left[{j_1\over \a ^2}+({k_s\over k_T})^{1/2}j_2\right]
\label{f1}
\ee
after obvious substitutions, with the two integrals 

\be
j_1=\int_{k_s\over k_T}^{\a^ 4k_s\over k_T}dy\, y^{1/12}e^{-y}H[({k_B\over k_s})^3-y]
\label{f2}
\ee
and 

\be
j_2=\int_{\a^ 4k_s\over k_T}^\infty dy\, y^{-5/12}e^{-y}H[({k_B\over k_s})^3-y]H[\gamma_B-k_s]
\label{f3}
\ee
Examining the Heaviside functions we obtain for $k_s<\gamma_B<\gamma _0$ 

\be
j_1(k_s<\gamma_B)=\int_{k_s\over k_T}^{\a^ 4k_s\over k_T}dy\, y^{1/12}e^{-y}
\label{f4}
\ee
and 

\be
j_2(k_s<\gamma_B)=\int_{\a^ 4k_s\over k_T}^{({k_B\over k_s})^3} dy\, y^{-5/12}e^{-y}
\label{f5}
\ee
Alternatively, for $\gamma_B\le k_s\le \gamma _0$ we find $j_2(k_s\ge \gamma_B)=0$ and 

\be
j_1(k_s\ge \gamma_B)=\int_{k_s\over k_T}^{({k_B\over k_s})^3}dy\, y^{1/12}e^{-y}
\label{f6}
\ee

In order to derive asymptotic fluence distribution we have to consider again limiting cases of 
the initial Thomson limit ($K=4k_0\gamma _0\ll 1$) and Klein-Nishina limit ($K\gg 1$). The Klein-Nishina limit each separate into two cases, 
depending on the value of $\a $ for $K \gg 1$:

\noi (a) the initial Thomson limit ($K\ll 1$), 

\noi (b) the mild initial Klein-Nishina limit ($1\ll K\ll \a ^3$), where $\gamma_B\ll k_B\ll \gamma _0\ll k_T$, and 

\noi (c) the extreme initial Klein-Nishina limit ($K\gg \a ^3 \gg 1$), where $k_B\ll \gamma_B\ll \gamma _0\ll k_T$. 

We consider each case in turn. 
\subsubsection{Initial Thomson limit}
In the initial Thomson limit the integration limit $\a^4k_s/k_T$ is small compared to unity for $k_s<k_T/\a ^4$, so that we obtain 
approximately 

\bdm
j_1(k_s<k_T/\a^4)\simeq {12\over 13}\a ^{13/3}({k_s\over k_T})^{13/12}
\edm
and 

\bdm 
j_2(k_s<k_T/\a^4)\simeq {12\over 7},
\edm
yielding for the bracket in Eq. (\ref{f1}) 

\be
{j_1\over \a ^2}+({k_s\over k_T})^{1/2}j_2\simeq {12\over 7}({k_s\over k_T})^{1/2}\left[1+{7\over 13}({\a ^4k_s\over k_T})^{7/12}\right]
\simeq {12\over 7}({k_s\over k_T})^{1/2}
\label{f7}
\ee
For scattered photon energies $(k_T/\a ^4)<k_s<k_T$ the integration limit $\a^4k_s/k_T$ is large compared to unity, so that according to 
Eq. (\ref{f5}) $j_2$ is exponentially small if $k_T<\gamma_B$ or vanishes if $k_T>\gamma_B$. However, $j_1\simeq 12/13$ in this scattered photon
energy range. Consequently, the bracket in Eq. (\ref{f1}) here is 

\be
{j_1\over \a ^2}+({k_s\over k_T})^{1/2}j_2\simeq {12\over 13\a ^2}
\label{f8}
\ee
For $k_s>k_T$ the dominating term to the bracket in Eq. (\ref{f1}) is 

\be
{j_1\over \a ^2}\simeq \a ^{2}({k_s\over k_T})^{13/12}e^{-k_s/k_T}
\label{f9}
\ee
Combining the asymptotics (\ref{f7})--(\ref{f9}) we find 

\be
F_{SSC}(k_s, K\ll 1\ll \a)\simeq {3S_0Q_0^2R^2k_0\over 7D_0\gamma _0^3}[{{k_T\over \a^4}\over {k_T\over \a ^4}+k_s}]^{1/2}
({k_s\over k_T})^{-1/4}e^{-k_s/k_T},
\label{f10}
\ee
which at energies above $k_T/\a ^4$ agrees with { Eq. (S-116) of paper S}. \\
{ We observe that, similarly to the \sy peak for $\a\gg 1$, the value of $\a$ determines the range of the power-law between the break $k_T/\a^4$ and the cut-off $k_T$.}
\subsubsection{Mild initial Klein-Nishina limit}
Here the integration boundary 

\be
{\a^4k_s\over k_T}={k_s\over (K\gamma_B/\a ^3)}
\label{f11}
\ee
is small (large) compared to unity for $k_s<(K\gamma_B/\a ^3)$ and $k_s>(K\gamma_B/\a ^3)$, { respectively}, because $K/\a ^3\ll 1$. 
For $k_s<(K\gamma_B/\a ^3)$ we obtain again the approximation (\ref{f7}) for the bracket in Eq. (\ref{f1}), whereas for 
$(K\gamma_B/\a ^3)<k_s<k_B$ the approximation (\ref{f8}) holds. 

For $\gamma_B\ll k_B<k_s\le \gamma _0\ll k_T$, $j_2=0$ and both integration limits in Eq. (\ref{f6}) are small compared to unity yielding 

\be
j_1\simeq {12\over 13}({k_B\over k_S})^{13/4}\left[1-({k_s\over \gamma _0})^{13/3}\right]
\label{f12}
\ee
We find 

\bdm
F_{SSC}(k_s, 1\ll K\ll \a^3)\simeq {3S_0Q_0^2R^2k_0\over 13D_0\gamma _0^3}
\edm
\be
\times
\cases{{13\over 7}\left({k_s\over k_T}\right)^{-1/4}  & \hbox{for} $k_s \ll (K\gamma_B/\a ^3)$, \cr 
{K\over \a ^2}\left({k_s\over k_B}\right)^{-3/4}  & \hbox{for} $(K\gamma_B/\a ^3)\ll k_s \ll k_B$,\cr
{K\over \a ^2}\left({k_s\over k_B}\right)^{-4}\left[1-({k_s\over \gamma _0})^{13/3}\right]  & \hbox{for} $k_B\ll k_s \le \gamma _0 $\cr }
\label{f13}
\ee
In the mild initial Klein-Nishina limit { of} the high injection case $\a \gg 1$ the SSC fluence exhibits a triple power law behavior with two spectral breaks
at $(K\gamma_B/\a ^3)$ and $\gamma_B$, { indicating again the possibility to determine $\a$ from the spectrum}. At high energies $k_s>(K\gamma_B/\a ^3)$ the fluence (\ref{f13}) agrees with { Eq. (S-119) of paper S}.
\subsubsection{Extreme initial Klein-Nishina limit}
Here the integration boundary (\ref{f11}) is small compared to unity for all photon energies $k_s\le \gamma_B$ because $K/\a ^3\gg 1$. 
For $k_s<k_B$ the approximation (\ref{f7}) for the bracket in Eq. (\ref{f1}) holds in this photon energy range, whereas for $k_s>\gamma_B$, $j_2=0$ 
and approximation (\ref{f12}) is valid. In the intermediate photon energy range $k_b<k_s<\gamma_B$ 
we obtain for the bracket in Eq. (\ref{f1})

\bdm
{j_1\over \a ^2}+({k_s\over k_T})^{1/2}j_2\simeq {12\over 13}\a ^{7/3}({k_s\over k_T})^{13/12}+
{12\over 7}{k_B^{7/4}\over k_T^{1/2}k_s^{5/4}}[1-({k_s\over \gamma_B})^{7/3}]
\edm
\be
\simeq{12\over 7}K^{-2/3}({k_s\over k_B})^{-5/4}
\label{f14}
\ee
We obtain 

\bdm
F_{SSC}(k_s, 1\ll \a^3\ll K)\simeq {3S_0Q_0^2R^2k_0\over 13D_0\gamma _0^3}
\edm
\be
\times
\cases{{169\over 84}\left(k_s/k_T\right)^{-1/4}  & \hbox{for} $k_s \ll k_B$,\cr 
{13\over 7}K^{1/3}(k_s/k_B)^{-2}    & \hbox{for} $k_B\ll k_s\ll \gamma_B$ \cr 
{K\over \a ^2}\left(k_s/k_B\right)^{-4}\left[1-({k_s\over \gamma _0})^{13/3}\right]  & \hbox{for} $\gamma_B\ll k_s \le \gamma _0$ \cr }
\label{f15}
\ee
a different triple power law with two spectral breaks at $k_B$ and $\gamma_B$, { hinting once more at the possibility to infer $\a$ from the spectrum}.
%
%
\section{Synchrotron and SSC SEDs} \label{sec5}
Measured data is normally displayed in spectral energy distribution plots (SED), which is the fluence multiplied with the frequency, or $f(\nu)=\nu F(\nu)$ in units of$\ergcm$. 

Transforming into the stationary frame of the galaxy, which is for low red-shifts approximately the observe's frame from earth, gives $\Dp(\nup{})=\delta^4 f(\nup{}/\delta)$, where primed quantities are measured in the stationary frame, and $\delta=[\Gamma_b(1-\beta_{\Gamma}\cos{\theta_{obs}})]^{-1}$ is the Doppler-factor, { with the Lorentz factor of the plasma blob $\Gamma_b$, the corresponding speed in units of $c$ of the blob $\beta_{b}=v_{b}/c=(1-\Gamma_b^{-2})^{1/2}$, and the angle between the jet and the line of sight $\theta_{obs}$. 

In many papers the dominance of one peak over the other is referred to as the Compton dominance, which is defined as the ratio of the peak values of the respective components in the SED: 
\be
R^{\prime}_{max}=\frac{\Dp_{SSC,max}}{\Dp_{s,max}}\equiv \frac{\Dp_{SSC}(\nup{}_{SSC,max})}{\Dp_{s}(\nup{}_{s,max})}.
\label{g0}
\ee
Here $\nup{}_{SSC,max}$ and $\nup{}_{s,max}$ refer to the peak frequencies of the SSC- and the \sy peak, respectively. We will also calculate the ratio of the peak frequencies, defined by 
\be
R^{\prime}_{peak} = \frac{\nup{}_{SSC,max}}{\nup{}_{s,max}}.
\label{g01}
\ee}

From now on we drop the subscript $s$ of the scattered SSC frequency, since the measured quantity is always $\nup{}$. 
\subsection{Small injection case $\a \ll 1$} \label{sec51}
For the low-injection case $\a \ll 1$ the \sy SED can be calculated from Eq. (\ref{b2}), giving
\be
\Dp_s(\nup{}) = 5.9 \cdot 10^{12} \delta^4 \frac{\a^2}{\gamma_4} \left( \frac{\nup{}}{\nu_{syn}} \right)^{1/2} e^{-\nup{}/\nu_{syn}} \ergcm ,
\label{g1}
\ee
with $\nu_{syn}=mc^2k_0\delta/h=4.1\cdot 10^{14}\delta b\gamma_4^2\us$. \\
The maximum value of the \sy SED, 
\be
\Dp_{s,max}=2.5\cdot 10^{12} \delta^4 \frac{\a^2}{\gamma_4} \ergcm,
\label{g2}
\ee
is attained at
\be
\nup{}_{s,max}=\frac{1}{2}\nus = 2.1\cdot 10^{14}\delta b\gamma_4^2 \us. 
\label{g3}
\ee
Likewise, Eqs. (\ref{e3}) and (\ref{e6}) yield for the SSC SEDs
\be
\Dp_{SSC}(K\ll 1) = 3.0\cdot 10^{13} \delta^4 \frac{\a^4}{\gamma_4} \left( \frac{\nup{}}{\nut} \right)^{3/4} e^{-\nup{}/\nut} \ergcm
\label{g4}
\ee
in the Thomson limit, while for the Klein-Nishina limit 
\be
\Dp_{SSC}(K\gg 1) = 2.2\cdot 10^{14} \delta^4 \frac{\a^4}{b\gamma_4^4} \frac{\left( \frac{\nup{}}{\nub} \right)^{3/4}}{\left( 1+\frac{\nup{}}{\nub} \right)^{7/4}} \left[ 1-\left( \frac{\nup{}}{\nug} \right)^{7/12} \right] \ergcm,
\label{g5}
\ee
with the constants $\nut=mc^2k_T\delta/h=1.6\cdot 10^{23}\delta b\gamma_4^4\us$, the Schlickeiser-R\"oken break frequency $\nub=mc^2k_B\delta/h=2.3\cdot 10^{24}\delta b^{-1/3}\us$, and $\nug=mc^2\gamma_0\delta/h=1.2\cdot 10^{24}\delta\gamma_4\us$. \\
The maximum frequencies
\be
\nup{}_{SSC,max}(K\ll 1) = \frac{3}{4}\nut = 1.2\cdot 10^{23} \delta b\gamma_4^4 \us
\label{g6}
\ee
and
\be
\nup{}_{SSC,max}(K\gg 1) = \frac{3}{4}\nub = 1.7\cdot 10^{24} \delta b^{-1/3} \us
\label{g7}
\ee
imply the maximum values
\be
\Dp_{SSC,max}(\Kl) = 1.1\cdot 10^{13} \delta^4 \frac{\a^4}{\gamma_4} \ergcm 
\label{g8}
\ee
and
\be
\Dp_{SSC,max}(\Kg) = 6.7\cdot 10^{13} \delta^4 \frac{\a^4}{b\gamma_4^4} \ergcm ,
\label{g9}
\ee
respectively. \\
\begin{figure}
	\centering
		\includegraphics[width=1.00\textwidth]{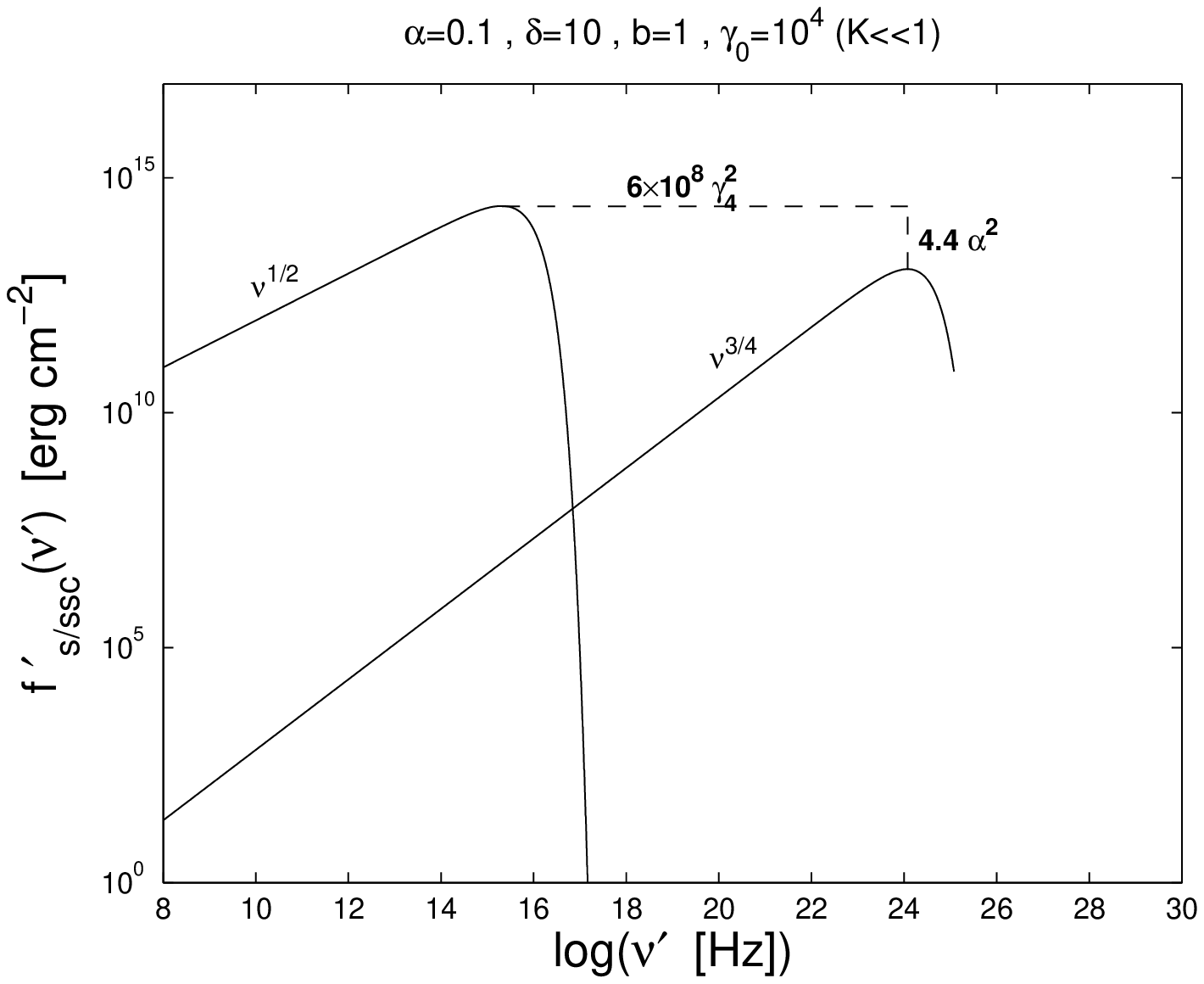}
	\caption{$\Dp$ as a function of $\nup{}$ for $\a\ll 1$ and $\Kl$. The parameters are indicated at the top. Displayed are the Compton dominance ($\Rpm{\Kl}$, vertical line), and the ratio of the peak frequencies ($\Rpp{\Kl}$, horizontal lines), as well as the power-laws.}
	\label{fig2}
\end{figure}

In { Fig. \ref{fig2}} we show the SED for the case of $\a\ll 1$ and $\Kl$. We set the free parameters to $\a=0.1$, $\delta=10$, $b=1$, and $\gamma_0=10^4$. { We indicate the power-laws, the Compton dominance} (vertical line)
\be
\Rpm{\Kl} = \frac{\Dp_{SSC,max}(\Kl)}{\Dp_{s,max}} = 4.4\a^2 ,
\label{g10}
\ee
and the ratio of the peak frequencies (horizontal line)
\be
\Rpp{\Kl} = \frac{\nup{}_{SSC,max}(K\ll 1)}{\nup{}_{s,max}} = 6\cdot 10^8 \gamma_4^2 .
\label{g11}
\ee
\begin{figure}
	\centering
		\includegraphics[width=1.00\textwidth]{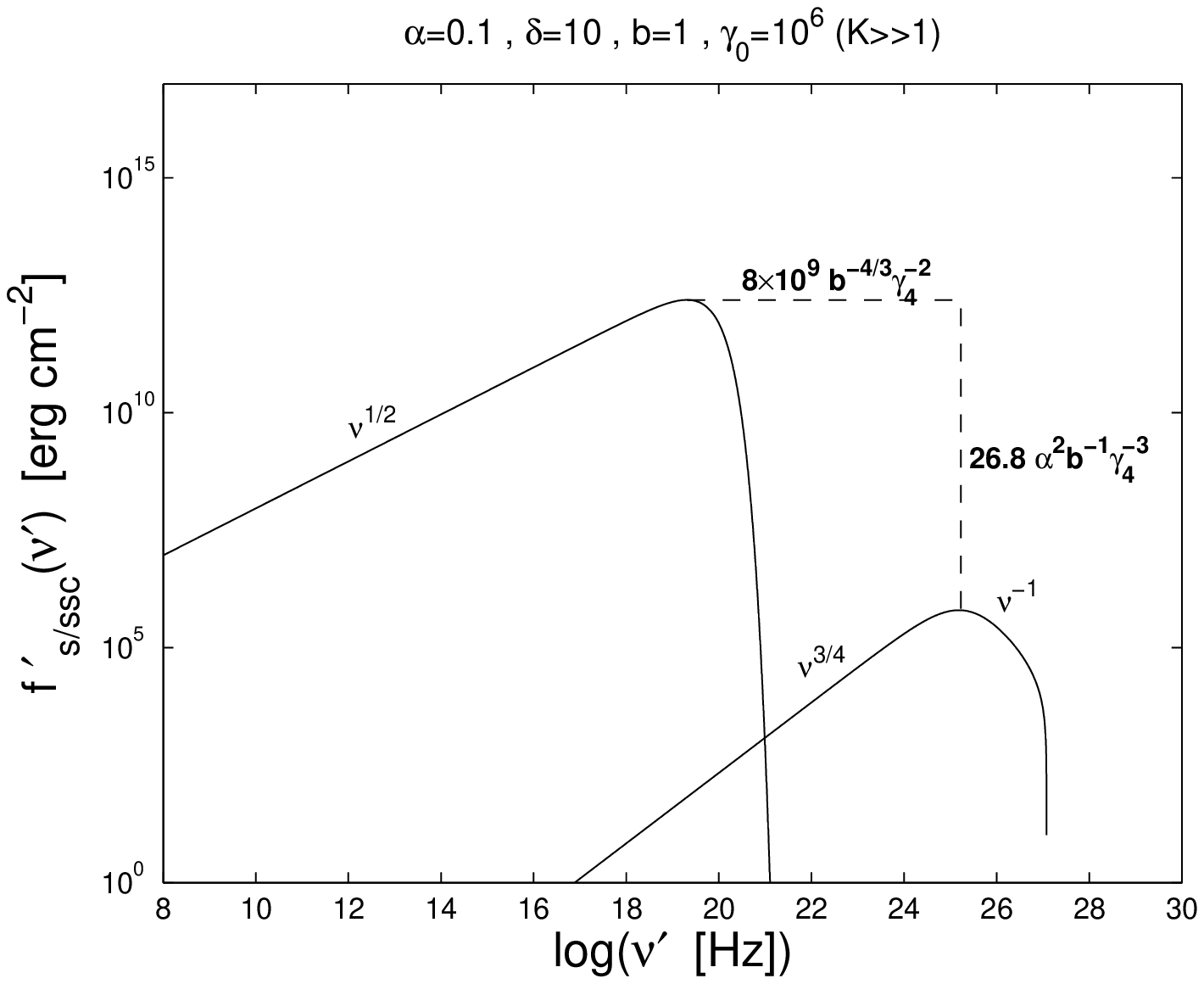}
	\caption{$\Dp$ as a function of $\nup{}$ for $\a\ll 1$ and $\Kg$. The parameters are indicated at the top. Displayed are the Compton dominance ($\Rpm{\Kg}$, vertical line), and the ratio of the peak frequencies ($\Rpp{\Kg}$, horizontal lines), as well as the power-laws.}
	\label{fig4}
\end{figure}

{ Fig. \ref{fig4} shows} the SED for the case of $\a\ll 1$ and $\Kg$. Apart from the initial Lorentz factor, which is now $\gamma_0=10^6$, we changed none of the free parameters { compared to Fig. \ref{fig2}. As before, we display the power-laws, as well as the Compton dominance} (vertical line)
\be
\Rpm{\Kg} = \frac{\Dp_{SSC,max}(\Kg)}{\Dp_{s,max}} = 26.8 \frac{\a^2}{b\gamma_4^3} ,
\label{g12}
\ee
and the ratio of the peak frequencies (horizontal line)
\be
\Rpp{\Kg} = \frac{\nup{}_{SSC,max}(\Kg)}{\nup{}_{s,max}} = 8\cdot 10^9 \frac{1}{b^{4/3}\gamma_4^2} .
\label{g13}
\ee
\subsection{Large injection case $\a \gg 1$}
Similarly to the case of a small injection parameter we will now discuss the SEDs for $\a\gg 1$. We begin with the synchrotron SED, which can be derived from Eq. (\ref{b4}):
\be
\Dp_{s} = 5.9\cdot 10^{12} \delta^4 \frac{\a^2}{\gamma_4} \frac{\left( \frac{\nup{}}{\nus} \right)^{1/2}}{1+\frac{\nup{}}{\nuc}} e^{-\nup{}/\nus} \ergcm .
\label{g14}
\ee
It peaks at
\be
\nup{}_{s,max} = \nuc = 2.9\cdot 10^{14} \delta \frac{b\gamma_4^2}{\a^2} \us
\label{g15}
\ee
with the maximum value
\be
\Dp_{s,max} = 2.5\cdot 10^{12} \delta^4 \frac{\a}{\gamma_4} \ergcm .
\label{g16}
\ee
For the SSC SEDs we have to discuss three different cases, and we begin with the Thomson limit $\Kl$ derived from Eq. (\ref{f10}):
\be
\Dp_{SSC}(\Kl) = 3.0\cdot 10^{13} \delta^{4} \frac{\a^4}{\gamma_4} \frac{\left( \frac{\nup{}}{\nut} \right)^{3/4}}{\left( 1 + \frac{\a^4\nup{}}{\nut} \right)^{1/2}} e^{-\nup{}/\nut} \ergcm .
\label{g17}
\ee
The maximum value 
\be
\Dp_{SSC,max}(\Kl) = 1.7\cdot 10^{13} \delta^4 \frac{\a^2}{b\gamma_4} \ergcm
\label{g18}
\ee
is attained at
\be
\nup{}_{SSC,max}(\Kl) = \nut = 4.0\cdot 10^{22} \delta b \gamma_4^4 \us .
\label{g19}
\ee
For the mild Klein-Nishina-limit $(\Kla)$ Eq. (\ref{f13}) is the point to begin with, resulting in
\bdm
\Dp_{SSC}(\Kla) = 
\edm
\be
\cases{3.0\cdot 10^{13} \delta^4 \frac{\a^4}{\gamma_4^4} \left( \frac{\nup{}}{\nut} \right)^{3/4} \ergcm  & \hbox{for} $\nup{} \ll \nuta$ \cr 
3.1\cdot 10^{13} \delta^{4} \frac{\a^2}{b^{1/3}\gamma_4^2} \frac{\left( \frac{\nup{}}{\nub} \right)^{1/4}}{\left( 1+\frac{\nup{}}{\nub} \right)^{13/4}} \left[ 1-\left( \frac{\nup{}}{\nug} \right)^{13/3} \right] \ergcm & \hbox{for} $\nup{} \gg \nuta $ .}
\label{g20}
\ee
This triple power-law peaks at
\be
\nup{}_{SSC,max}(\Kla) = \frac{1}{12}\nub = 1.9\cdot 10^{23} \delta b^{-1/3} \us
\label{g21}
\ee
and reaches a maximum value of
\be
\Dp_{SSC,max}(\Kla) = 1.5\cdot 10^{13} \delta^4 b^{-1/3} \frac{\a^2}{\gamma_4^2} \ergcm .
\label{g22}
\ee
The last case is the extreme Klein-Nishina limit $(\Kga)$. We obtain the SED from Eq. (\ref{f15}) yielding
\bdm
\Dp_{SSC}(\Kga) = 
\edm
\be
\cases{2.2\cdot 10^{14} \delta^4 \frac{\a^4}{b\gamma_4^4} \frac{ \left( \frac{\nup{}}{\nub} \right)^{3/4}}{\left( 1+\frac{\nup{}}{\nub} \right)^{7/4}} \ergcm & \hbox{for} $\nup{} \ll \nuga$ \cr 
3.1\cdot 10^{13} \delta^4 \frac{\a^2}{b^{1/3}\gamma_4^2} \left( \frac{\nup{}}{\nub} \right)^{-3} \left[ 1-\left( \frac{\nup{}}{\nug} \right)^{13/3} \right] \ergcm  & \hbox{for} $\nup{} \gg \nuga $ .}
\label{g23}
\ee
The SED peaks with a maximum value
\be
\Dp_{SSC,max}(\Kga) = 5.5\cdot 10^{13} \delta^4 b^{-1/3} \frac{\a^2}{\gamma_4^2} \ergcm
\label{g24}
\ee
and a peak frequency
\be
\nup{}_{SSC,max}(\Kga) = \frac{3}{4}\nub = 1.7\cdot 10^{24} \delta b^{-1/3} \us .
\label{g25}
\ee
As in section \ref{sec51} we plot the respective SEDs { with the indication of the power-laws and the respective ratios.} 
\begin{figure}
	\centering
		\includegraphics[width=1.00\textwidth]{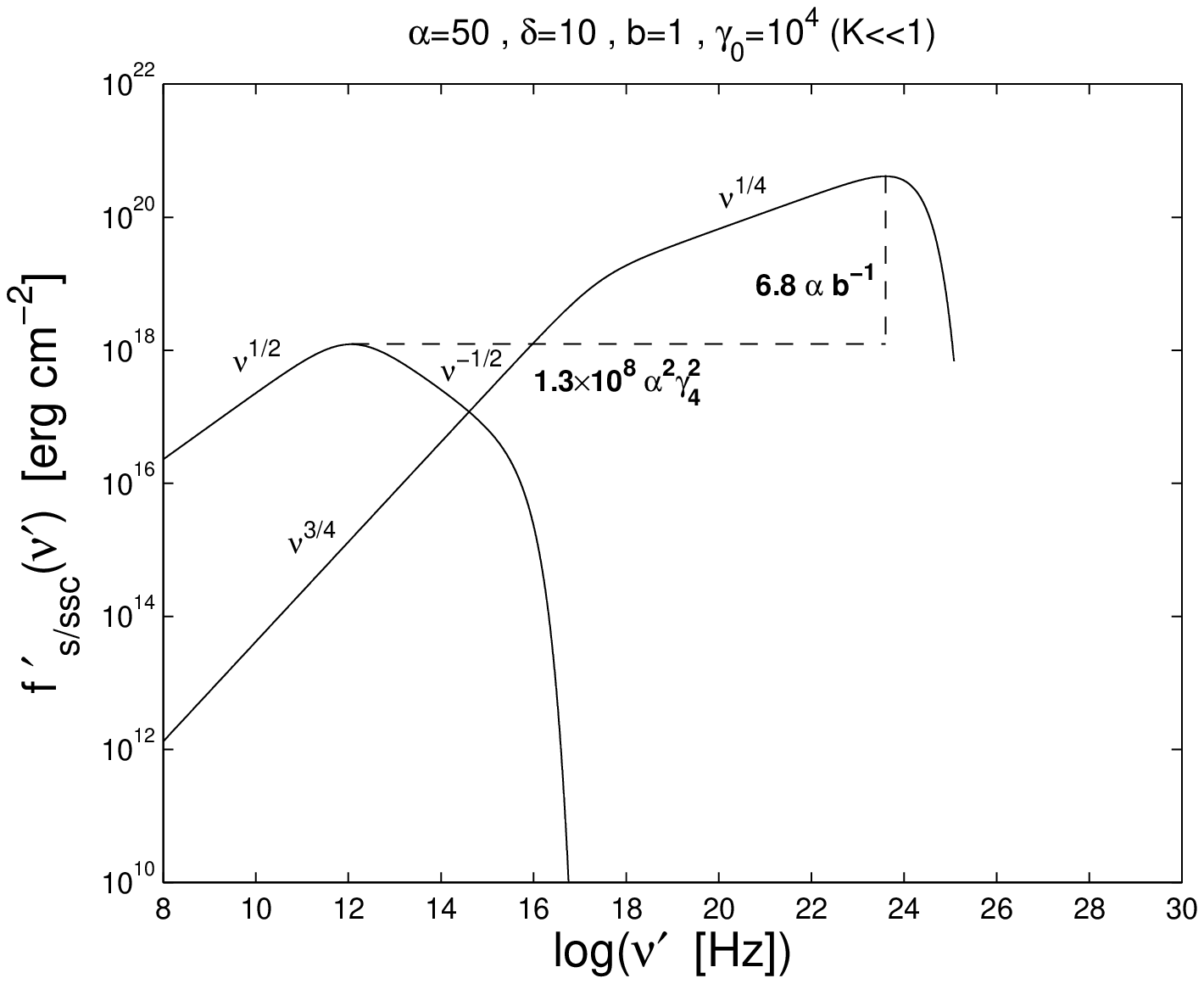}
	\caption{$\Dp$ as a function of $\nup{}$ for $\a\gg 1$ and $\Kl$. The parameters are indicated at the top. Displayed are the Compton dominance ($\Rpm{\Kl}$, vertical line), and the ratio of the peak frequencies ($\Rpp{\Kl}$, horizontal lines), as well as the power-laws.}
	\label{fig6}
\end{figure}

{ Fig. \ref{fig6} presents} the case of $\a\gg 1$ and $\Kl$, with the specific parameters $\a=50$, $\delta=10$, $b=1$, and $\gamma_0=10^4$. The Compton dominance becomes
\be
\Rpm{\Kl} = \frac{\Dp_{SSC,max}(\Kl)}{\Dp_{s,max}} = 6.8 \frac{\a}{b} ,
\label{g26}
\ee
while the ratio of the peak frequencies 
\be
\Rpp{\Kl} = \frac{\nup{}_{SSC,max}(\Kl)}{\nup{}_{s,max}} = 1.3\cdot 10^{8} \a^2\gamma_4^2 .
\label{g27}
\ee
\begin{figure}
	\centering
		\includegraphics[width=1.00\textwidth]{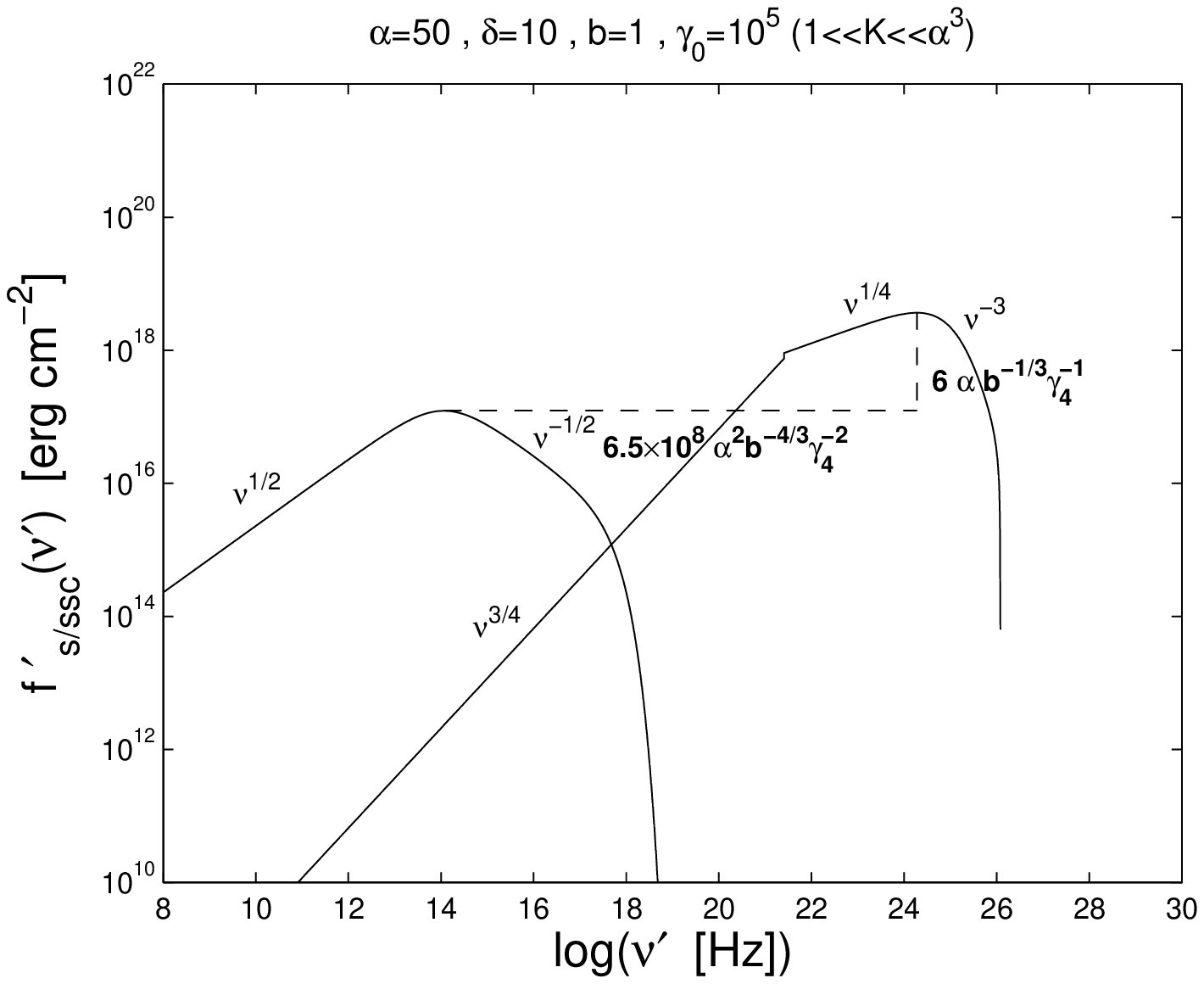}
	\caption{$\Dp$ as a function of $\nup{}$ for $\a\gg 1$ and $\Kla$. The parameters are indicated at the top. Displayed are the Compton dominance ($\Rpm{\Kla}$, vertical line), and the ratio of the peak frequencies ($\Rpp{\Kla}$, horizontal lines), as well as the power-laws.}
	\label{fig8}
\end{figure}

In order to display the mild Klein-Nishina case in { Fig. \ref{fig8}} we changed the value of the initial Lorentz-factor to $\gamma_0=10^5$, while the other parameters remain the same. The Compton dominance yields
\be
\Rpm{\Kla} = \frac{\Dp_{SSC,max}(\Kla)}{\Dp_{s,max}} = 6.0 \frac{\a}{b^{1/3}\gamma_4} ,
\label{g28}
\ee
and the ratio of the peak frequencies is in this case
\be
\Rpp{\Kla} = \frac{\nup{}_{SSC,max}(\Kla)}{\nup{}_{s,max}} = 6.5\cdot 10^{8} \frac{\a^2}{b^{4/3}\gamma_4^2} .
\label{g29}
\ee
\begin{figure}
	\centering
		\includegraphics[width=1.00\textwidth]{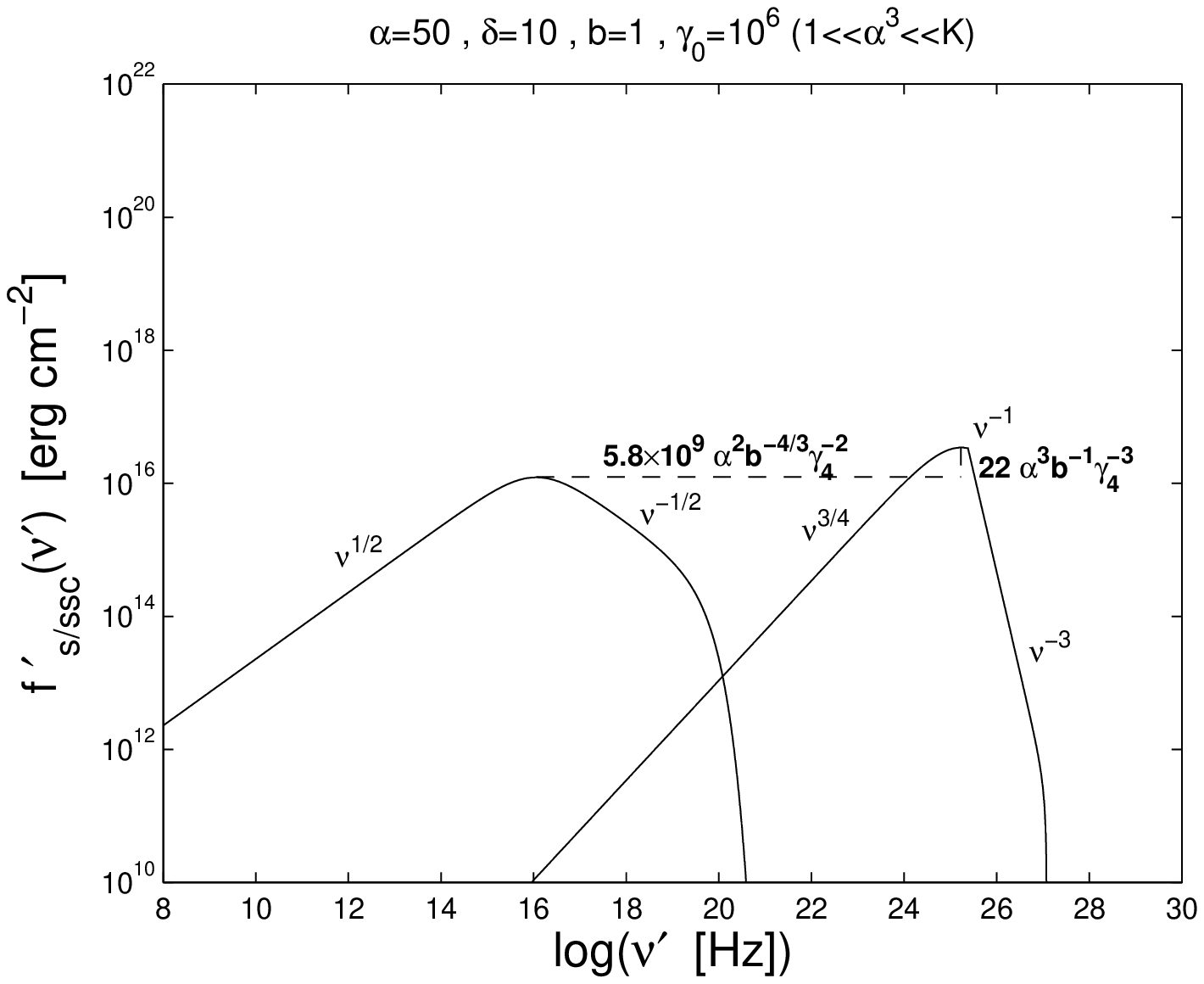}
	\caption{$\Dp$ as a function of $\nup{}$ for $\a\gg 1$ and $\Kga$. The parameters are indicated at the top. Displayed are the Compton dominance ($\Rpm{\Kga}$, vertical line), and the ratio of the peak frequencies ($\Rpp{\Kga}$, horizontal lines), as well as the power-laws.}
	\label{fig10}
\end{figure}

Finally, in { Fig. \ref{fig10}} we show the extreme Klein-Nishina case, which is obtained by $\gamma_0=10^6$. The other parameters are, again, unchanged { compared} to the previous two cases. We obtain for the Compton dominance and the peak frequencies 
\be
\Rpm{\Kga} = \frac{\Dp_{SSC,max}(\Kga)}{\Dp_{s,max}} = 22.0 \frac{\a^3}{b\gamma_4^3} ,
\label{g30}
\ee
and 
\be
\Rpp{\Kga} = \frac{\nup{}_{SSC,max}(\Kga)}{\nup{}_{s,max}} = 5.8\cdot 10^{9} \frac{\a^2}{b^{4/3}\gamma_4^2} ,
\label{g31}
\ee
respectively. 
\subsection{Discussion}
Let us briefly discuss the above results, which show some remarkable points. 

First of all, the Compton dominance is clearly dependent on $\a$. For $\a\ll 1$ the inverse Compton peak is less luminous than the \sy peak, while in the opposite case the inverse Compton peak is dominant. This is what we expected and proves a posteriori our assumption that the luminosities of the peaks are directly related to the cooling terms and reflect the initial dominance of either the linear or the non-linear cooling. 

Another obvious result is that the synchrotron peak differs in the respective cases of $\a$. For $\a\ll 1$ it shows the { well-known} behaviour with a single power-law $\nup{1/2}$ followed by an exponential cut-off. In the other case the peak exhibits a broken power-law, first rising as usual with $\nup{1/2}$, but then decreasing with $\nup{-1/2}$ before it eventually cuts off exponentially. This broken power-law is a unique feature of the initial non-linear cooling, and we stress that it can give observers a clear indication if this non-linear process is at work in a blazar provided the dominance of the inverse Compton peak. 

Regarding the inverse Compton peak, we note that its luminosity decreases strongly with increasing $\gamma_0$. This is expected, since for larger $\gamma_0$ the Thomson cross section of the photon-electron collisions is replaced by the Klein-Nishina cross section, which is much reduced compared to the former. Still, for $\a\gg 1$ even in the extreme Klein-Nishina limit the inverse Compton peak dominates the synchrotron peak, while for $\a\ll 1$ the latter exhibits higher fluxes. 

{ An important observation is that for $\a\gg 1$ the width of the second power-law depends on $\a$ in both the synchrotron and the inverse Compton peak. One can, therefore, conclude that there is a direct relation between the $\nup{-1/2}$ power-law of the synchrotron peak and the $\nup{1/4}$ ($\nup{-1}$ in the extreme Klein-Nishina limit) power-law of the inverse Compton peak. Of course, the same is true for the $\nup{1/2}$ power-law of the synchrotron peak and the $\nup{3/4}$ power-law of the inverse Compton peak. This has an important consequence: 

As we showed in ZS the $\nup{-1/2}$ feature of the synchrotron peak for $\a\gg 1$ is unrelated to the injected electron distribution. It appears at the same interval in the spectrum for power-law electrons as well as for monoenergetic electrons. Only for higher energetic photons the injection distribution becomes important. We argue, now, that this should also be the case for the inverse Compton peak, since we showed above that parts of the peaks are related with each other. Thus, any injection distribution that differs from our monoenergetic approach here should only reveal itself after the second break in the inverse Compton peak, which is the maximum of the peak in case of the Thomson and the mild Klein-Nishina limit. Below that break the spectrum should have the unique features we outlined above.} 

We also note that the cut-off of the inverse Compton peak reflects the initial Lorentz factor $\gamma_0$, and gives a clear indication which limit needs to be taken into account. 

%
%
\section{Illustrative examples}
{ In this section we intend to compare our analytical results from section \ref{sec5} with flares of the sources 3C 279 and 3C 454.3. For that purpose we extracted the SEDs of the sources from the respective papers (see below), determined the rough location of the peaks, and calculated the free parameters. The resulting theoretical curves are overlaid in the plots of the SEDs.

Since we only try to illustrate the capability of our model, this is merely a fit "by eye", and we did not make any statistical analyses.

Although we already performed the transformation from the blob system to the system of rest in section \ref{sec5}, we did not yet account for distance effects diminishing the source. We, therefore, introduce another factor into the equations for $\Dp$ and $\Dp_{max}$, which we name $N_{orm}$ combining all necessary effects. This may seem arbitrary, since $N_{orm}$ could contain a lot of free parameters. However, it mostly depends on the size of the blob and the luminosity distance of the source. The latter is a known quantity, as long as the red-shift of the source is determined. In that cases, $N_{orm}$ would depend only on the size of the blob.

Thus, we have five free parameters, which we cannot determine altogether, since we only have four values from the peaks of the sources. We could try to get another value from a break, but this would be even more uncertain, and so we leave one parameter open, from which we calculate the others with the help of the equations given in section \ref{sec5}.}

\subsection{The case of 3C 279}
\begin{figure}
	\centering
		\includegraphics[width=1.00\textwidth]{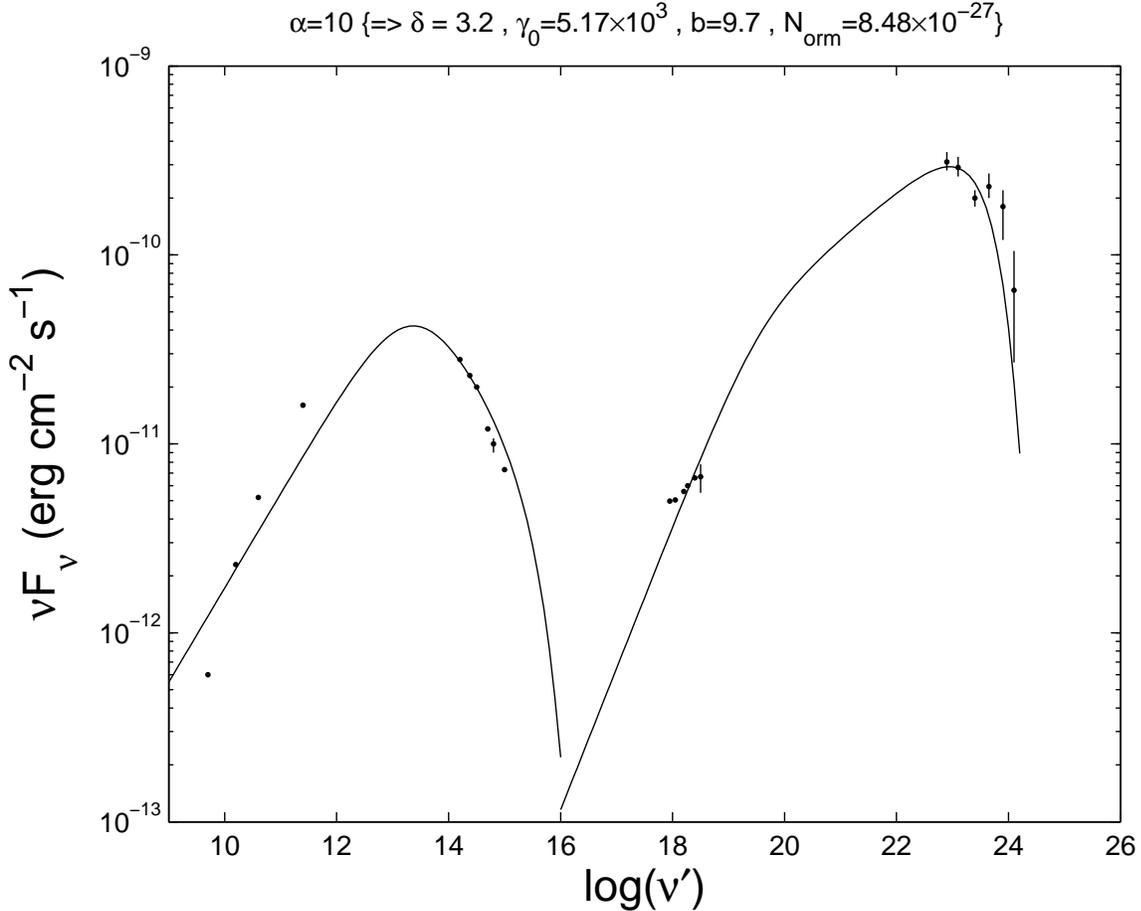}
	\caption{The SED of the outburst of 3C 279 in February 2009 (data extracted from Abdo et al. (2010)) overlaid with our model from Eq. (\ref{g14}) and (\ref{g17}). The model parameters are indicated at the top. $\a$ is the remaining free parameter, leading to the values of the other ones.}
	\label{fig11}
\end{figure}

Abdo et al. (2010) reported about a powerful outburst of the blazar 3C 279 in February 2009. The SED of this outburst (Fig. \ref{fig11}; { red dots in Fig. 2 of Abdo et al. (2010)}) shows that the inverse Compton peak dominates the \sy peak, { leading to the case $\a\gg 1$}.  

{ The location and maximum values of the \sy and SSC peaks are determined as $\nup{}_{s,max}\simeq 10^{13.4}\us$ and $\Dp_{s,max}\simeq 4.3\times 10^{-11} \ergcm s^{-1}$, as well as $\nup{}_{SSC,max}\simeq 10^{22.9}\us$ and $\Dp_{SSC,max}\simeq 3.0\times 10^{-10} \ergcm s^{-1}$, respectively. This readily gives us the Compton dominance $R^{\prime}_{max}\approx 7.5$, and the ratio of the peak frequencies $R^{\prime}_{peak}\approx 3.3\cdot 10^{9}$. As mentioned in the previous section, the cut-off of the inverse Compton peaks gives some direct hints of the initial Lorentz factor $\gamma_0$, and, thusly, which limit we need to take into account. Since the cut-off is located at a frequency of about $10^{24}\us$,  we conclude that the source operates in the Thomson limit. With the help of Eqs. (\ref{g15}), (\ref{g16}), (\ref{g18}), and (\ref{g19}) (although any two equations can be replaced by Eqs. (\ref{g26}), and (\ref{g27})), we can calculate four of the five free parameters.

Setting $\a=10$ results in $\delta=3.2$, $\gamma_0=5.17\times 10^3$, $b=9.7$, and $N_{orm}=8.48\times 10^{-27}$.} The emerging curve is also plotted in Fig. \ref{fig11}. 

Obviously, we achieve a rather good fit with our parameters. The \sy peak is matched well at least after the maximum, { where we can recover the slope of the spectrum without the need for fancy electron distributions}. The increasing part of that peak is not covered as good. However, we can argue that the lowest data point belongs to a regime where the blob becomes optically thick for \sy photons, an effect we did not account for in our model.

We also achieve a very good fit for the inverse Compton peak. Even the highest energies are matched rather well. { The underrepresentation of the highest energies could be due to our monoenergetic approach, leading to a premature cut-off. A broader electron distribution might cover the highest energies with better accuracy, while the dip in the spectrum can be due to $\gamma-\gamma$ attenuation.} 

The parameters we used for our model curve are also quite reasonable. The Doppler-factor of only $3.2$ is rather low and far less than values estimated from other models which needed Doppler factors of almost $100$ to model blazar spectra, { while most observations hint at small Doppler factors (Henri \& Sauge 2006)}. Previous solutions invoked a more complicated structure of the jet, like the jet-in-a-jet model by Giannios et al. (2009). The initial Lorentz factor of the electrons is also rather low, and should be easily achievable by acceleration processes like Fermi I and II, or magnetic reconnection events. The magnetic field strength of $9.1$ G seems higher than what is usually invoked (see e.g. Ghisellini et al. 2009), but it is still within an order of magnitude of most values, { and far below the values needed in hadronic models (e.g. Zacharopoulou et al. 2011)}. 

\subsection{The case of 3C 454.3}
\begin{figure}
	\centering
		\includegraphics[width=1.00\textwidth]{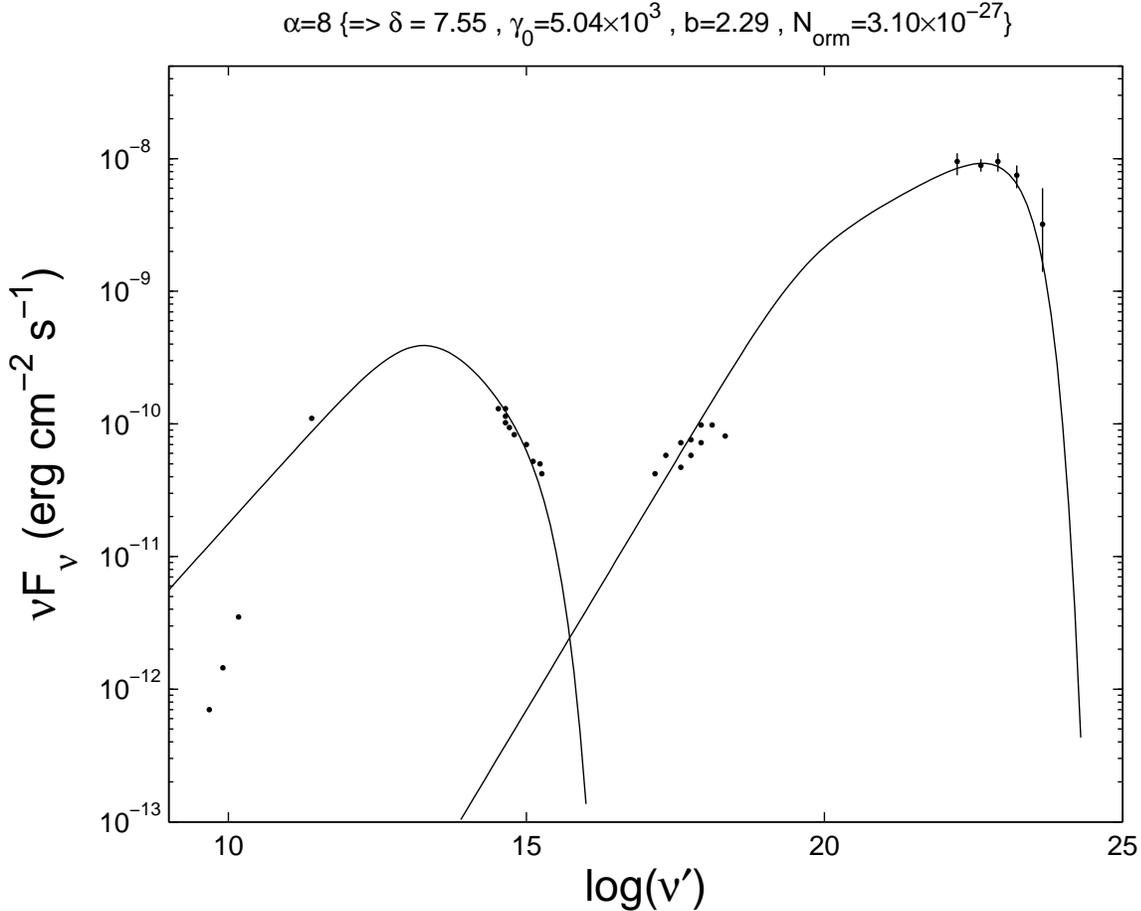}
	\caption{The SED of the outburst of 3C 454.3 in November 2010 (data extracted from Vercellone et al. (2011)) overlaid with our model from Eq. (\ref{g14}) and (\ref{g17}). The model parameters are indicated at the top. $\a$ is the remaining free parameter, leading to the values of the other ones.}
	\label{fig12}
\end{figure}

{ 3C 454.3 is reported as one of the most active blazars, showing several $\gamma$-ray outburst in the last few years. Vercellone et al. (2011) report of an exceptional flare in November 2010, where the source exhibited its highest state ever recorded reaching a flux six times higher than the Vela pulsar. Vercellone et al. (2011) presented three different SEDs, with one covering the outburst data, while the other two are from  before and after the flare, respectively. 

We extracted the data of the outburst SED (red filled circles in Fig. 4 of Vercellone et al. (2011)) and present it in Fig. \ref{fig12}. Obviously, as in the case of 3C 279, the inverse Compton peak dominates the \sy peak, which was interpreted as due to external Compton scattering by Vercellone et al. (2011). We try to fit our model to the data, for which we use the case $\a\gg 1$ in the Thomson limit.}

{ The peak frequencies and peak values are $\nup{}_{s,max}\simeq 10^{13.3}\us$ and $\Dp_{s,max}\simeq 4.0\times 10^{-10} \ergcm s^{-1}$, as well as $\nup{}_{SSC,max}\simeq 10^{22.6}\us$ and $\Dp_{SSC,max}\simeq 9.5\times 10^{-9} \ergcm s^{-1}$, respectively. This results in a Compton dominance of $R^{\prime}_{max}\approx 23.75$, and a ratio of the peak frequencies of $R^{\prime}_{peak}\approx 2.11\cdot 10^{9}$. As before, we are now able to calculate four of five parameters.

Choosing $\a=8$ as the remaining free parameters, we obtain $\delta=7.55$, $\gamma_0=5.04\times 10^3$, $b=2.29$, and $N_{orm}=3.10\times 10^{-27}$. The resulting curve is also plotted in Fig. \ref{fig12}.

We achieve, again, a very good fit of the data. In principle, the same arguments hold here, as well, that we used for 3C 279, namely that we neglected the optically thick regime for low \sy frequencies, and the underrepresentation of the highest $\gamma$-energies due to the monoenergetic cut-off. A possible dip in the $\gamma$-ray data due to $\gamma-\gamma$ attenuation is, at least, less obvious than for 3C 279.

The parameters are also quite reasonable, again, with the same arguments as above. While the Doppler factor is a little higher than in 3C 279, the magnetic field strength is little lower, while the initial Lorentz factor of the injected electrons is more or less the same.}

%
\section{Summary and conclusions}
We investigated the radiative signatures of blazars under the influence of the linear synchrotron cooling in combination with the non-linear synchrotron-self Compton cooling of a monoenergetic distribution of electrons. { Expanding the previous work of SBM, who calculated the electron distribution and the \sy spectrum, we determined the emerging SSC spectrum. This enabled us to give a complete prediction of the SED of blazar flares under the influence of the combined cooling term. Since the cooling term is time-dependent, the usual steady-state approach could not be followed. Instead, SBM solved the kinetic equation self-consistently. Such self-consistent analytical calculations of the complete SED have not been done so far, and the results are quite remarkable.} 

We found that the emerging spectra depend critically on the initial { ratio} of the cooling terms, which we described by the { injection} parameter $\a$, and also on the initial electron Lorentz-factor $\gamma_0$. 

For $\a\ll 1$ the cooling is completely linear and the well-known results of the SED are obtained. The synchrotron peak shows a single power-law $\nup{1/2}$ followed by an exponential cut-off. 

The SSC peak depends on the initial electron energy, which determines if the inverse Compton collisions take place in the Thomson or Klein-Nishina limit. For the former case the SSC SED also exhibits a single power-law $\nup{3/4}$ and an exponential cut-off. The latter case shows a broken power-law, where the power-law below the peak is the same as in the Thomson limit, while for higher photon energies the SED decreases as $\nup{-1}$ before it is also cut off. The luminosity of the SSC peak in the Klein-Nishina limit is reduced compared to the Thomson limit, which is not surprising, since the cross section of the Klein-Nishina limit is also much smaller than the one in the Thomson limit. 

We also found, in accordance with our expectations, that for $\a\ll 1$ the \sy peak dominates the inverse Compton peak. 

For $\a\gg 1$ the cooling is initially nonlinear and only becomes linear after some time. This has some remarkable consequences for the resulting SEDs. Especially the \sy peak exhibits a unique feature, which is only due to the non-linear cooling and which has also been obtained for an injection scenario including a power-law of electrons. Namely, the \sy spectrum increases first with $\nup{1/2}$ as in the low-$\a$ case, but after the maximum it decreases with $\nup{-1/2}$, and is only then followed by an exponential cut-off. 

The inverse Compton peak also shows a different appearance compared to the previous cases. In the Thomson limit its broken power-law with $\nup{3/4}$ and $\nup{1/4}$ is eventually cut off exponentially after the maximum. The Klein-Nishina limit is divided into two separated cases by an interplay of $\a$ and $\gamma_0$. In both cases the spectrum contains a triple power-law followed by a cut-off, which is, however, not exponential. The mild case includes $\nup{3/4}$, $\nup{1/4}$, and $\nup{-3}$ { power-laws}, while the extreme case contains $\nup{3/4}$, $\nup{-1}$, and $\nup{-3}$ { power-laws}. Although these power-laws look similar, they are clearly distinguished by the position of the peaks and the breaks. Also the luminosity of the inverse Compton peak declines from limit to limit. 

Nonetheless, according to our expectations, the inverse Compton peak dominates the \sy peak in every limit. This is remarkable, since most calculations and modeling attempts using the usual approach show the inverse Compton peak to be less luminous than the \sy peak and need external photon sources to account for a dominating inverse Compton peak. 

The { injection} parameter $\a$, which we first defined as the ratio of the cooling terms determining the initial dominance of one of them, is directly related to the { so-called Compton dominance, given by the ratio of the peak values in the SED. It reflects the initial condition of the source, being directly related to the energy and the density of the injected electrons. Therefore, with the help of $\a$, sources can be ordered according to their internal initial conditions. In our model a Compton dominance larger than unity is due to a large density of electrons in the plasma blob, and not necessarily due to external radiation fields. A low density of electrons, on the other hand, yields a Compton dominance smaller than unity.}

In order to apply our model to actual sources, we combined our results with the data of exceptional flares of the blazars 3C 279 in February 2009 (Abdo et al. 2010), { and 3C 454.3 in November 2010 (Vercellone et al. 2011). In both cases,} we can fit the data quite well with the case $\a\gg 1$ in the Thomson limit. The parameters {for 3C 279 are} $\a=10$, $\delta=3.2$, $b=9.7$, and $\gamma_0=5.17\times 10^3$, { while for 3C 454.3 we obtain $\a=8$, $\delta=7.55$, $b=2.29$, and $\gamma_0=5.04\times 10^3$. These values} are very reasonable, and indicate the potential of our approach. 

In order to make even more convincing statements more work needs to be done, especially more modeling attempts and also predictions and comparisons of light curves. The latter are important for variability analyses. We showed in ZS that the cooling of the electrons begins much faster after injection for the non-linear process compared to the linear one. This hints towards a shorter variability timescale, but needs to be examined in greater detail in the future. 

{ Similarly, the effects of \sy self-absorption and $\gamma-\gamma$ attenuation need to be taken into account, since especially a large $\a$ implies a large electron density in the plasma blob, leading to possibly strong photon absorptions. Currently, we deal with these problems, trying to obtain an even more realistic model.}

Finally, we stress once more that a dominating inverse Compton peak and a broken power-law of the \sy peak in the form given above hint clearly towards the non-linear SST cooling process being at work { during flaring states} in blazars. 
\section*{Acknowledgements}
We thank the anonymous referee for the constructive comments, which helped significantly improving the manuscript. \\
We acknowledge support from the German Ministry for Education and Research (BMBF) through Verbundforschung Astroteilchenphysik grant 05A08PC1 and the Deutsche Forschungsgemeinschaft through grants Schl 201/20-1 and Schl 201/23-1.

\end{document}